\documentclass{emulateapj}

\slugcomment{Accepted version from October 15, 2018}
\shorttitle{Tidal gas streams around NGC\,4631}
\shortauthors{P. Richter et al.}

\begin{document}

\title{
Circumgalactic gas at its extreme:\\
tidal gas streams around the Whale galaxy
NGC\,4631 explored with HST/COS
}

\author{
P. Richter\altaffilmark{1},
B. Winkel\altaffilmark{2},
B.~P. Wakker\altaffilmark{3},
N.~M. Pingel\altaffilmark{4,5}
A.~J. Fox\altaffilmark{6},
G. Heald\altaffilmark{7},
R.~A.~M. Walterbos\altaffilmark{8},
C. Fechner\altaffilmark{1},
N. Ben\,Bekhti\altaffilmark{9},
G. Gentile\altaffilmark{10},
L. Zschaechner\altaffilmark{11,12}
\vspace{0.3cm}\\
}

\affil{$^1$Institut f\"ur Physik und Astronomie, Universit\"at Potsdam,
Karl-Liebknecht-Str.\,24/25, 14476 Golm, Germany}
\affil{$^2$Max-Planck-Institut f\"ur Radioastronomie (MPIfR), Auf dem H\"ugel 69,
53121 Bonn, Germany}
\affil{$^3$Supported by NASA/NSF, affiliated with the Department of Astronomy,
University of Wisconsin-Madison,\\ 475 North Charter Street, Madison, WI 53706, USA}
\affil{$^{4}$Department of Physics and Astronomy, West Virginia University, White Hall,
Box 6315, Morgantown, WV 26506, USA}
\affil{$^{5}$Center for Gravitational Waves and Cosmology, West Virginia University,\\
Chestnut Ridge Research Building, Morgantown, WV 26505, USA}
\affil{$^6$Space Telescope Science Institute, 3700 San Martin Drive,
Baltimore, MD 21218, USA}
\affil{$^7$CSIRO Astronomy and Space Science, PO Box 1130, Bentley WA 6102, Australia}
\affil{$^{8}$Department of Astronomy, New Mexico State University, P.O. Box 30001, MSC 4500,
Las Cruces, NM 88003, USA}
\affil{$^9$Fraunhofer Institute for High Frequency Physics and Radar Techniques FHR,
Fraunhoferstr. 20, 53343 Wachtberg, Germany}
\affil{$^{10}$Department of Physics and Astrophysics, Vrije Universiteit Brussel, 
Pleinlaan 2, 1050 Brussels, Belgium}
\affil{$^{11}$University of Helsinki, Physicum, Helsingin Yliopisto,
Gustaf H\"allstr\"omin katu 2, 00560 Helsinki, Finland}
\affil{$^{12}$Finnish Center for Astronomy with ESO}

%

\begin{abstract}

We present a detailed analysis of the absorption properties
of one of the tidal gas streams around the ``Whale'' galaxy NGC\,4631
in the direction of the quasar 2MASS\,J12421031$+$3214268. Our study
is based on ultraviolet spectral data obtained with
the Cosmic Origins Spectrograph (COS) onboard the
{\it Hubble Space Telescope} (HST) and 21cm-data from the HALOGAS project
and the Green Bank Telescope (GBT).
We detect strong  H\,{\sc i} Ly\,$\alpha$ absorption in the velocity range 
$+550$ to $+800$ km\,s$^{-1}$ related to gas from a NGC\,4631 tidal stream 
known as Spur 2. We measure a column density of 
log ($N$(H\,{\sc i}/cm$^{-2}))=18.68\pm 0.15$, indicating that
the quasar sightline traces the outer boundary of Spur 2 as seen in the 21cm data.
Metal absorption in Spur 2 is detected in the lines of O\,{\sc i}, C\,{\sc ii}, 
Si\,{\sc ii}, and Si\,{\sc iii} in a complex absorption pattern that reflects
the multi-phase nature of the gas. We find that the average neutral gas fraction
in Spur 2 towards 2MASS\,J12421031$+$3214268 is only $14$ percent. This 
implies that ionized gas dominates the total mass of Spur 2, which then may comprise
more than $10^{9} M_{\sun}$. No significant depletion of Si is observed, showing
that Spur 2 does not contain significant amounts of dust.
From the measured O\,{\sc i}/H\,{\sc i} column-density ratio we determine an 
$\alpha$ abundance in Spur 2 of $0.13^{+0.07}_{-0.05}$ solar 
([$\alpha$/H$]=-0.90\pm 0.16$), which is substantially lower than what is observed
in the NGC\,4631 disk. The low metallicity and low dust content 
suggest that Spur 2 represents metal-deficient gas stripped off a gas-rich 
satellite galaxy during a recent encounter with NGC\,4631.
\vspace{0.5cm}\\
\end{abstract}

%


%

\section{Introduction}

Galaxy evolution at low and high redshift is believed to be
largely influenced by the galaxies' interactions with
their immediate cosmological environment. As most galaxies are
expected to consume their interstellar gas content on relatively
short timescales to sustain star formation, the inflow of gas
from outside belongs to the particularly interesting processes that
are believed to strongly influence a galaxy's star-formation rate
and its evolutionary state (e.g., Dav\'e et al.\,2012).
Indeed, recent observational and theoretical studies have
demonstrated that the inflow of gaseous material from the
intergalactic medium (IGM) and the recycling of previously expelled
gas adds significantly to the today's reservoir of baryonic matter in a
galaxy, thus founding the basis for future star-formation activity
(see recent reviews by Kacprzak 2017, S\'anchez-Almeida 2017, Finlator 2017,
and references therein).

In addition to IGM accretion and gas recycling, the gravitational interaction
of galaxies (including minor and major mergers) represents yet another
important mechanism by which galaxies acquire gas. Merger processes can
deposit huge amounts of
gaseous material in the form of tidal streams into the circumgalactic region
around interacting galaxies galaxies long before the main stellar bodies of
the galaxies are in close proximity (Fraternali \& Binney 2008;
Di\,Teodoro \& Fraternali 2014). And because the gravitational interaction
also pulls out gas from the inner-most regions of the merging galaxies, tidal
gas streams often contain large amounts of neutral gas, visible at low redshift
in the 21cm emission line of neutral hydrogen and (at low and high redshift) in
damped absorption in the Ly\,$\alpha$ line at $1215.7$ \AA.

The most prominent local example for a massive tidal gas stream is the
Magellanic Stream in the halo of the Milky Way, which originates
in the gravitational and hydrodynamical interaction between 
the Magellanic Clouds and the Milky Way
(e.g., Gardiner \& Noguchi 1996; Connors et al.\,2006;
Besla et al.\,2010, 2012; Nidever et al.\,2010; Diaz \& Bekki 2011, 2012;
D'Onghia \& Fox 2016; Salem et al.\,2015).
Recent absorption-line observations (Fox et al.\,2013, 2014, 2018; Richter
et al.\,2013, 2018) and emission-line observations (Barger et al.\,2013) 
indicate that the Magellanic Stream,
when considered together with the Leading Arm and the Magellanic Bridge
between the Large and the Small Magellanic Cloud,
probably contains more than
$3\times 10^9 M_{\sun}$ of (mostly ionized) gas, thus exceeding the combined
interstellar gas mass of the two Magellanic Clouds (Br\"uns et al.\,2005;
Fox et al.\,2014; Richter 2017).
Other prominent examples for extended tidal gas streams visible in 21cm
emission can be seen in the M81 group (e.g., Yun et al.\,1994;
Chynoweth et al.\,2008) and in the ``Whale'' galaxy NGC\,4631
(Rand 1994), the latter being the subject of the study presented here.

Detailed case studies of the circumgalactic medium (CGM) in nearby, merging
galaxies hold the prospect of improving our understanding of the role of
galaxy interactions for the distribution and physical properties of
circumgalactic gas in the general context of galaxy evolution.
To reliably estimate the chemical composition of the CGM,
its distribution, its kinematics, and its total mass, the multi-phase
nature of the gas needs to be taken into account. For this, the
combination of deep 21cm observations (tracing neutral hydrogen gas
and its spatial distribution) and high signal-to-noise (S/N) ultraviolet (UV)
absorption-line measurements (tracing metal absorption in the neutral
and ionized circumgalactic gas phases) offers a particularly powerful
and efficient observing strategy.

In this paper, we present new UV absorption-line measurements of circumgalactic
gas belonging to the tidal gas streams around the Whale galaxy NGC\,4631
in the direction of the background quasar 2MASS\,J12421031$+$3214268. The
quasar is located at an impact parameter of $\rho=40$ kpc to NGC\,4631,
a value that is comparable to the size of the galaxy's H\,{\sc i} disk
(Fig.\,1).
From the analysis of the UV data we obtain new information on
the metal and dust abundance in the south-western part of the gaseous
streams, its origin, and its total mass.

Our paper is organized as follows: in Sect.\,2 we describe the overall
properties of the tidal
gas streams around NGC\,4631 and summarize previous measurements.
In Sect.\,3 we outline the HST/COS observations, the supplementary
HALOGAS and GBT 21cm observations, and the analysis methods.
The results from the absorption-line analysis are presented
in Sect.\,4. In Sect.\,5 we discuss the implications of our results
in regard to the origin and mass of the tidal gas
around NGC\,4631 and compare its properties with the
Magellanic Stream. Our study is summarized in Sect.\,6.


\section{Tidal gas streams around the Whale galaxy}

One particularly spectacular example for a tidal interaction
between galaxies of different masses is the Whale galaxy NGC\,4631
with its various companions (e.g., NGC\,4656 and NGC\,4627).
NGC\,4631 itself  is a nearby ($d=7.4\pm0.2$ Mpc; $cz=606$ km\,s$^{-1}$;
Radburn-Smith et al.\,2011)
edge-on spiral galaxy of Hubble type SB(s)d with a current
star-formation rate of $2.1$ $M_{\sun}$ yr$^{-1}$ (Kennicut et al.\,2008;
Sanders et al.\,2003). It
is located in a galaxy group environment containing
more than a dozen galaxies (Giuricin et al.\,2000).
From field imaging surveys it is known that there are plenty
of dwarf satellite galaxies around NGC 4631 (Ann et al.\,2011).
Because most of these satellite galaxies are believed
to be located within the virial radius
of NGC\,4631, there must be strong on-going tidal and hydrodynamical
interactions between NGC\,4631 and its satellite-galaxy population.
The larger companion galaxies NGC\,4627 and NGC\,4656 are located
at $2.6 \arcmin$ to the north-west and $32 \arcmin$ to the
south-east of NGC\,4631. Therefore, with an assumed
distance of 7.4 Mpc for NGC\,4631, NGC 4627 and NGC 4656
are separated by about 6 kpc and 71 kpc, respectively,
from NGC\,4631.

In particular NGC\,4656 is a probable candidate for a close
encounter which leads to tidal stripping of disk
material, because it is massive enough to detach gas
and stars from the inner regions of NGC\,4631. The distorted
morphology of the dwarf elliptical galaxy NGC\,4627 and
the well-known H\,{\sc i} bridge between NGC\,4631
and NGC\,4656 (Roberts 1968; Weliachew et al.\,1978; Rand 1994) are 
clear signs for the strong interactions between NGC\,4631 and its
satellite galaxies. A detailed study of the 21cm properties
of the gaseous streams around NGC\,4631 was presented by
Rand (1994), who grouped the various tidal features into
five major spurs (see Fig.\,1). Combes (1978) showed that a
parabolic encounter of NGC\,4656 with NGC\,4631 can
explain part of the circumgalactic H\,{\sc i} distribution (Spurs 1, 4 and 5).
Spurs 2 and 3, on the other hand, could represent material stripped from
a smaller companion of NGC\,4631 (e.g., NGC\,4627) or could
be stirred-up material from the NGC\,4631 disk from a close
encounter of NGC\,4631 with one of its satellites.
The latter scenario is supported by the fact that Spur 2
smoothly joins onto the outer disk of NGC\,4631 in the
21cm data cube (Rand 1994).

The widespread presence of circumgalactic gas components in NGC\,4631
is also indicated by a variety of other multi-wavelength
observations, from X-rays over H$\alpha$ to other radio wavebands
(Fabbiano et al.\,1987; Wang et al.\,1995; Neininger et al.\,1999;
Bendo et al.\,2006; Martin \& Kern 2001).
The detection of (hot) ionized gas far away from the galactic plane
was seen as evidence that NGC\,4631 drives a galactic
superwind (e.g., Melo et al.\,2002; Yamasaki 2009).

While the many previous observations of NGC\,4631 and
its environment already unveil a fascinating amount of
details on the many aspects that drive the evolution
of the galaxy in this complicated environment, an absorption-line
study of the CGM against one or several background QSOs
(essential to derive the metallicity of the gas and its total
mass) has not been carried out so far.
With this study, we are closing this gap.

%

\begin{figure*}[t!]
\epsscale{1.1}
\plotone{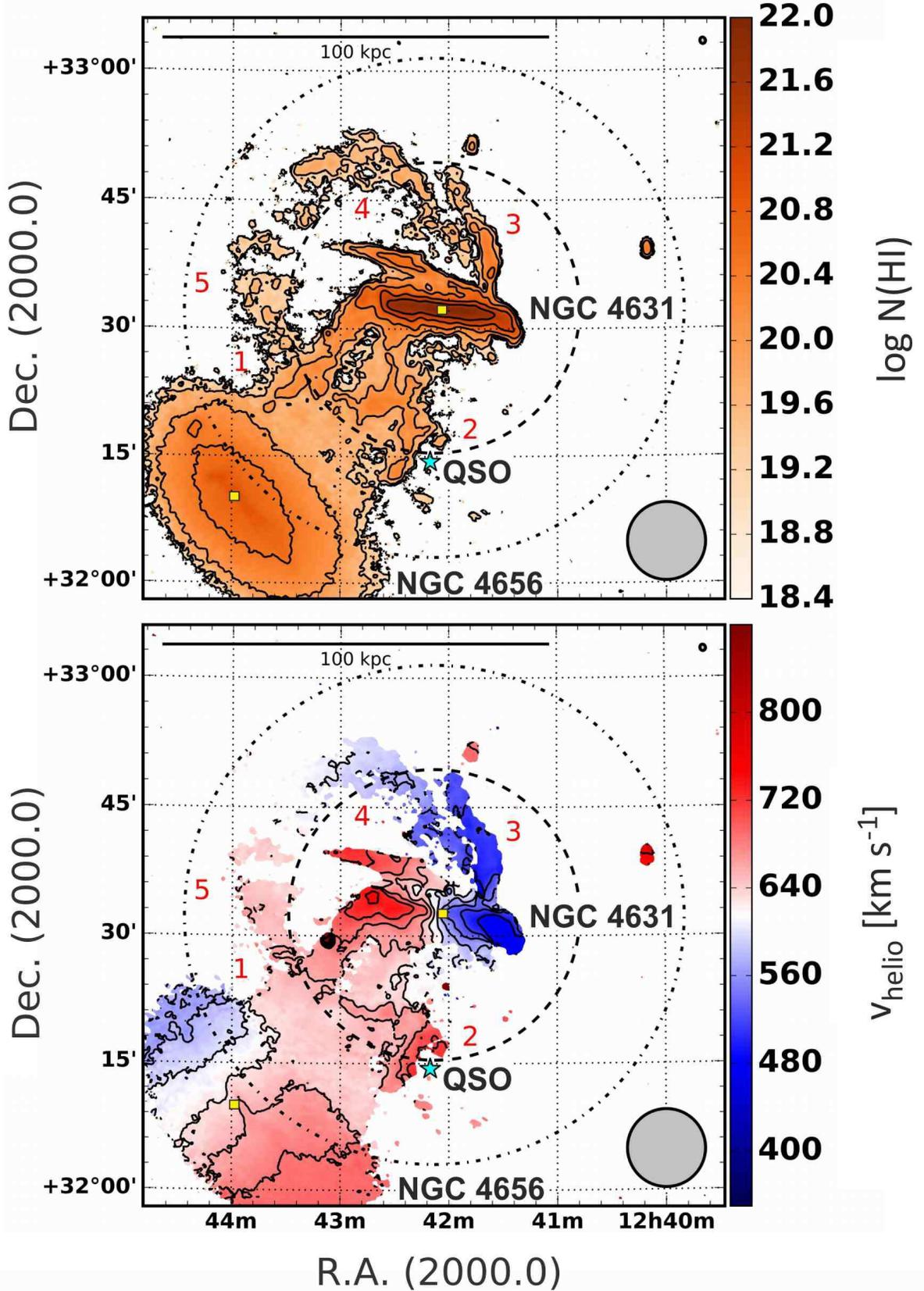}
\caption{
Spatial distribution of H\,{\sc i} 21cm emission
around NGC\,4631 and its neighboring galaxies
from the HALOGAS/GBT observations.
{\it Upper panel:} distribution of 21cm emission
with the H\,{\sc i} column density colour-coded.
The H\,{\sc i} column density contours range from
log $N$(H\,{\sc i}$)=18.5$ to $22$ in steps of $0.5$ dex.
{\it Lower panel:} distribution of 21cm emission
with the radial velocity colour-coded.
The velocity contours begin at $+350$ km\,s$^{-1}$, 
end at $+880$ km\,s$^{-1}$, and increase in increments of 
$20$ km\,s$^{-1}$.
The position of the background QSO
2MASS\,J12421031$+$3214268 is labeled
with the blue star symbol. The optical center positions of
NGC\,4631 and NGC\,4656 are indicated with the small yellow boxes. 
The various CGM components/spurs
are labeled following the scheme presented in Rand (1994).
The beamsizes appropriate for the field center and field edge
are shown as gray circles in the top-right and 
bottom-right corner, respectively.
The dashed and dot-dashed circles indicate the 50 percent 
and 10 percent levels of the WSRT primary beam response
(see Sect.\,3.2). 
}
\end{figure*}

%

\begin{figure*}[t!]
\epsscale{1.0}
\plotone{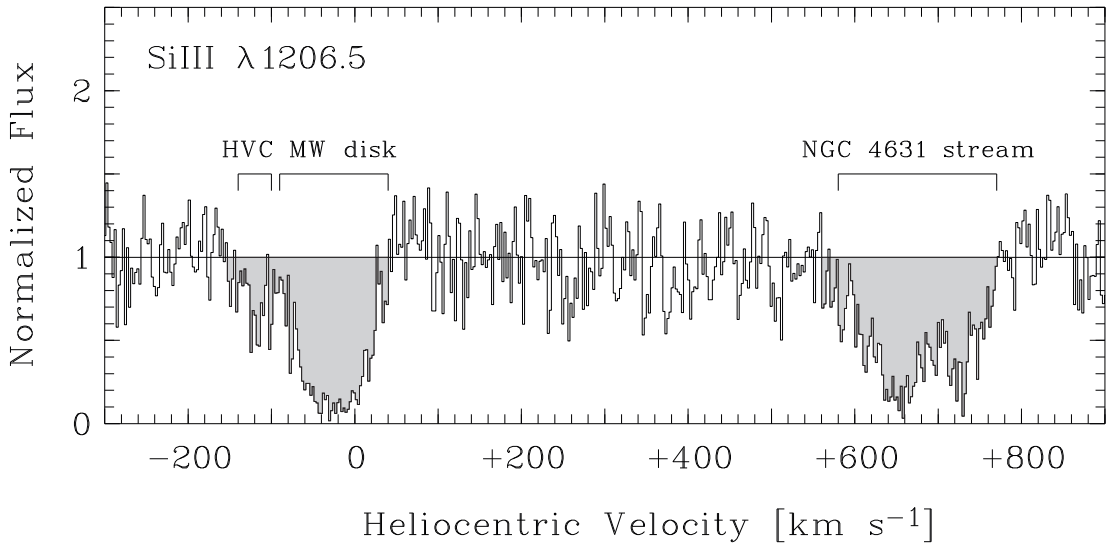}
\caption{
Velocity-component structure of Milky Way/NGC\,4631 absorption towards
2MASS\,J12421031$+$3214268 in the
Si\,{\sc iii} $\lambda 1206.5$ line (in the heliocentric velocity
frame). Milky Way absorption from the various disk and halo (high-velocity
cloud; HVC) components is seen between $-160$ and $+60$ km\,s$^{-1}$.
Absorption related to the NGC\,4631 tidal stream in front of the QSO is observed
in the range $v_{\rm helio}=+550$ and $800$ km\,s$^{-1}$.}
\end{figure*}

%

\begin{figure}[t!]
\epsscale{0.90}
\plotone{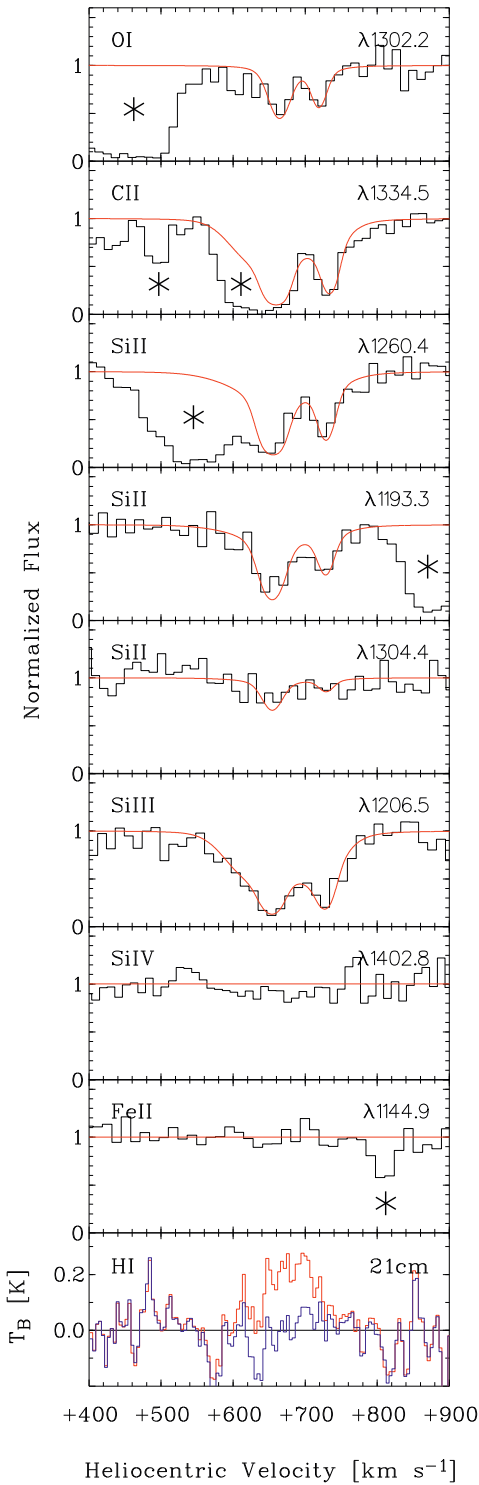}
\caption{
Absorption profiles of metal ions in the COS spectrum of 2MASS\,J12421031$+$3214268
and 21cm emission profiles from the WSRT/GBT data.
For displaying purposes, the COS data have been binned over 5 pixels.
Absorption from Spur 2 around NGC\,4631 is seen
in the velocity range $v_{\rm helio}=+550$ to $+800$ km\,s$^{-1}$.
The red solid line in the absorption spectra shows the best fitting absorber model, as described
in Sect.\,4.2. The star symbols indicate blending lines not
related to Spur 2. In the lowest panel (21cm data), the
blue and red solid lines indicate the emission spectra from the WSRT only and
WSRT+GBT combined data set in the direction of 2MASS\,J12421031$+$3214268, respectively.
}
\end{figure}


\section{Observations and data handling}

\subsection{HST/COS observations and spectral analysis}

HST/COS observations of the background QSO 2MASS\,J12421031$+$3214268
(alternative name: SDSS J124210.34+321427.2;
$\alpha_{2000}=12^{\rm h}42^{\rm m}10.32^{\rm s}$,
$\delta_{2000}=+32\deg 14^{\rm m}26.80^{\rm s}$,
$z_{\rm m}=1.49$,
$m_{\rm FUV}=18.78$)
were carried out on April 19, 2016 (proposal ID: 14085; PI: P. Richter).
The position of the QSO sightline with respect to the neutral gas
distribution around NGC\,4631 and its companion galaxies is
indicated in Fig.\,1. As can be seen, the 2MASS\,J12421031$+$3214268
sightline passes the outer boundaries of Spur 2 south of the NGC\,4631
disk.

For the COS observations we used the G130M grating (centered on $\lambda=1291.0$ \AA),
which covers a wavelength range between $1150$ and $1450$ \AA.
The spectral resolution at the COS lifetime position 3
is $R=18,000$, corresponding to a velocity resolution of
$17$ km\,s$^{-1}$ FWHM and
a native pixel size of $3$ km\,s$^{-1}$ (Green et al.\,2012; Debes et
al.\,2016). The data for 2MASS\,J12421031$+$3214268 were
collected over five HST orbits with a total integration time of
$13.6$ ks. The individual science exposures were processed with
the CALCOS data reduction pipeline v3.1.7 and then co-added.
For the co-addition we used a custom-written code that aligns individual
exposures based on a pixel/wavelength calibration that uses several
ISM anchor lines (see Richter et al.\,2016, 2017).

The COS G130M data cover the following strong ion transitions:
H\,{\sc i} $\lambda 1215.7$,
C\,{\sc ii} $\lambda 1334.5$,
N\,{\sc i} $\lambda 1200.7$,
O\,{\sc i} $\lambda 1302.2$,
Si\,{\sc ii} $\lambda\lambda 1190.4,1193.3,1260.4,1304.4$,
Si\,{\sc iii} $\lambda 1206.5$,
Si\,{\sc iv} $\lambda\lambda 1393.8,1402.8$,
P\,{\sc ii} $\lambda 1152.8$,
S\,{\sc ii}  $\lambda 1253.8$,
and Fe\,{\sc ii} $\lambda\lambda 1143.2,1144.9$.
These diagnostic ion transitions trace weakly and moderately (photo)ionized
gas at temperatures $T<10^5$ K.
Because the important O\,{\sc i} $\lambda 1302.2$ line generally is
contaminated by airglow lines in COS day-time data, we conducted
a special re-processing of our 2MASS\,J12421031$+$3214268 data set
with night-only exposure intervals that were taken from the photon list
of the individual exposures.

For the spectral analysis of the COS data we used the {\tt span} software package,
which is based on the {\tt fitlyman} routines implemented in ESO-MIDAS
(Fontana \& Ballester 1996). The {\tt span} program allows us to 
determine equivalent widths, column densities, and Doppler
parameters using either a direct pixel integration in combination
with the apparent-optical depth (AOD) method (Savage \& Sembach 1991), a
curve-of-growth method, or a Voigt-profile fitting/modeling method that
takes into account the wavelength-dependent line-spread function of the
spectrograph (here: COS). Wavelengths and oscillator strengths of the
analyzed ion transitions were adopted from the list of Morton (2003).

Note that another, bright QSO close to NGC\,4631
(SDSS J124044.68$+$330349.8) was also observed with COS as part of the
HST observing program 14085. Although the GALEX
UV magnitude is $m_{\rm FUV}=18.78$, the COS data of SDSS$-$J124044.68$+$330349.8
does not show any significant UV flux. Since the pointing was checked to be correct, this
flux deficit most likely is due to a  Lyman-limit system (LLS)
at higher redshift. As a consequence,
the COS spectrum of SDSS$-$J124044.68$+$330349.8 does not yield any
usable information for our study.


\subsection{Supplementary WSRT/HALOGAS and GBT 21cm data}

We supplement our HST/COS data with 21cm observations of NGC\,4631
from the Hydrogen Accretion in LOcal GalaxieS project
(HALOGAS, PI: G. Heald; Heald et al.\,2011). HALOGAS represents a
comprehensive study of nearby galaxies using the Westerbork Synthesis
Radio Telescope (WSRT). The initial goal was to perform
ultra-deep H\,{\sc i} observations of 24 nearby spiral galaxies to
search for neutral gas in the circumgalactic environment of these
galaxies (see Zschaechner et al.\,2011; Gentile et al.\,2013; Kamphuis et al.\,2013
for results on individual HALOGAS galaxies).
In the framework of HALOGAS, NGC\,4631 was observed with the WSRT for
$10\times 12\,\mathrm{hrs}$ in the ``maxishort'' configuration with
a characteristic beam size of $15 \arcsec$ and a velocity resolution
of $\sim 4$ km\,s$^{-1}$
(see Heald et al.\,2011 for more details on the observations).

Additional 21cm observations of the NGC\,4631 environment have been carried out 
with the Robert C. Byrd Green Bank Telescope\footnote{The Green Bank Observatory 
is a facility of the National Science Foundation, operated under a cooperative 
agreement by AssociatedUniversities, Inc.}, which provides an angular resolution 
of $9.1 \arcmin$ at 1420 MHz.
The GBT data were obtained during in 2014 as part of project GBT16B 293. 
An area of $2\times2$ deg$^2$ was mapped down to a theoretical column density 
level of $10^{18}$ cm$^{-2}$ (at the GBT resolution), 
which corresponds to a total mapping time of 10 hours.  
The observational techniques used are the same as described in Pisano (2014), while 
the calibration and imaging was done with the same script as in Pingel et al.\,(in prep).  
Less than $0.1$ percent of these data were flagged for RFI, then smoothed to 
$6.4$ km\,s$^{-1}$ wide velocity bins.

To combine the GBT with the WSRT data, the Common Astronomy Software Applications (CASA)
task ``feather'' was used. A detailed description on the combination procedure
is provided in the Appendix. 
The integrated total 21cm column density map and the velocity field of NGC\,4631 from the
combined WSRT/GBT observations are shown in Fig.\,1 with
the position of the QSO indicated with the blue star symbol.
The combined data set provides a spectacular new view on the complex
H\,{\sc i} distribution in the tidal gas features around NGC\,4631 and NGC\,4656,
complementing previous deep 21cm observations (Weliachev et al.\,1978; Rand 1994).
For the interpretation of Fig.\,1 it is important to note that the effective 
angular resolution of the combined dataset varies with location in the field of view.

As opposed to a mosaic, the original WSRT observations were performed with a single 
pointing. Because the standard reduction techniques of imaging and deconvolution of 
interferometer data result in a model representation of the sky multiplied by the 
primary-beam response of the antennas, the true sensitivity of the native WSRT data 
follows the response of the primary- beam and thus decreases towards the edge of 
the field of view. Over-plotting the 50 percent (dashed) and 10 percent (dot-dashed) 
level of the primary beam response (Fig.\,1) according to the numerical model 
discussed in Heald et al.\,(2011) reveals that the disk of NGC\,4656 is well outside 
the half-power response of the WSRT primary beam. Effectively, the sensitivity 
of the original WSRT data becomes significantly reduced towards the H\,{\sc i} 
associated with NGC\,4656. On the other hand, because the GBT data were obtained 
via multiple pointings over four square degrees, there is no variation in the 
sensitivity of the single dish data over the same field of view. As a result, 
the angular scale of the diffuse emission surrounding NGC\,4656 in the combined 
image is almost certainly dominated by the angular resolution of the GBT data. 
The apparent vertical extent of the diffuse halo encompassing NGC\,4646 of $\sim 20$ kpc 
in the combined image (Fig.\,1) is therefore likely an upper limit.

That said, the GBT data detect almost ten times more H\,{\sc i} (in terms of mass) 
than the native WSRT data, indicating there is extended structure that is 
both resolved out by the native WSRT data and missed due to the 
decreased sensitivity towards the companion. Additional deep 21cm 
interferometer data would be very helpful to explore the full extent of
diffuse 21cm emission in the larger environemnt of NGC\,4631 and NGC\,4656.


\begin{deluxetable}{llllll}
\tabletypesize{\scriptsize}
\tablewidth{0pt}
\tablecaption{Summary of column-density measurements}
\tablehead{
\colhead{Ion} & \colhead{$\lambda_0$ [\AA]} & \colhead{$f$\tablenotemark{a}} 
& \colhead{$W_{\lambda}$ [m\AA]} & \colhead{log $N_{\rm AOD}$\tablenotemark{b}} 
& \colhead{log $N_{\rm model}$\tablenotemark{c}}
} 
\startdata
O\,{\sc i}    & 1302.17     & 0.048 & $160 \pm 23$ & $14.49 \pm 0.09$ & $14.54 \pm 0.08$ \\
C\,{\sc ii}\tablenotemark{d} 
              & 1334.53     & 0.128 & $162-412$    & $\geq 14.11$     & $14.93^{+0.09}_{-0.53}$\\
Si\,{\sc ii}  & 1193.29     & 0.582 & $278 \pm 26$ & $13.74 \pm 0.11$ & $13.91 \pm 0.07$\\
              & 1260.42     & 1.176 & $387 \pm 44$ & $\geq 13.62$     & $13.91 \pm 0.07$\\
              & 1304.37     & 0.086 & $\leq 71$    & $\leq 13.94$     & $13.91 \pm 0.07$\\
Si\,{\sc iii} & 1206.50     & 1.627 & $488 \pm 32$ & $\geq 13.61$     & $13.79 \pm 0.12$\\
Si\,{\sc iv}  & 1402.77     & 0.254 & $\leq 67$    & $\leq 13.28$     & ...   \\
Fe\,{\sc ii}  & 1144.94     & 0.083 & $\leq 36$    & $\leq 13.79$     & ...   
\enddata
\tablenotetext{a}{Oscillator strength.}
\tablenotetext{b}{Total column density in Spur 2, obtained by integrating the ion profiles in
the range $v_{\rm helio}=550-800$ km\,s$^{-1}$.}
\tablenotetext{c}{Total column density in Spur 2, obtained by summing over $N$ in all velocity
components based on the component model described in Sect.\,4.2.}
\tablenotetext{d}{C\,{\sc ii} is blended by intervening H\,{\sc i} Ly\,$\zeta$ absorption
at $z=0.043677$.}
\end{deluxetable}


\begin{figure}[hb!]
\epsscale{1.0}
\plotone{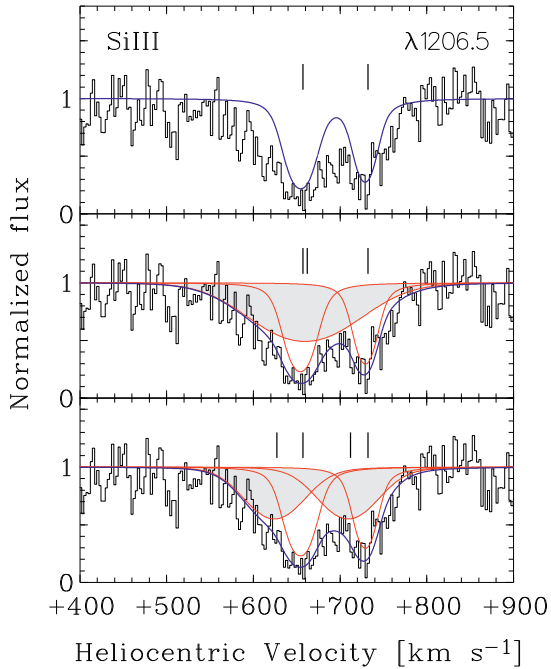}
\caption{
De-composition of absorption profiles in the COS data, here 
demonstrated for the strong Si\,{\sc iii} $\lambda 1206.5$ line. 
Next to the two narrow compontents 1 and 2 (upper panel)
a third, very broad satellite component (middle panel) or, alternatively,
two additional, moderately broadened satellite components (lower panel)
are required in the component model to match the data
(see Sect.\,4.2 for details).
}
\end{figure}

%

\section{Results}

\subsection{Metal absorption in the COS data}

Metal absorption towards 2MASS\,J12421031$+$3214268
from Spur 2 around NGC\,4631 is seen
in the velocity range $v_{\rm helio}=+550$ to $+800$
km\,s$^{-1}$ in various transitions from the metal ions
O\,{\sc i}, Si\,{\sc ii}, C\,{\sc ii}, and Si\,{\sc iii}.
Absorption from N\,{\sc i}, Si\,{\sc iv}, Fe\,{\sc ii},
and other ions is not detected.

The velocity-component structure of Milky Way plus NGC\,4631 absorption
towards 2MASS\,J12421031$+$3214268
between $-300$ and $+900$ km\,s$^{-1}$ is shown in Fig.\,2 for the
example of the Si\,{\sc iii} $\lambda 1206.5$ line. Milky Way (MW) disk
absorption is seen between $v_{\rm helio}=-90$ to $+50$ km\,s$^{-1}$,
while Galactic-halo absorption from a high-velocity cloud (HVC)
is seen near $-110$ km\,s$^{-1}$. The strong NGC\,4631 absorption at
$v_{\rm helio}=+550$ to $+800$ km\,s$^{-1}$ is broad 
and asymmetric, already indicating
the presence of multiple velocity sub-components in the gas.

In Fig.\,3 we display the velocity profiles of several lines from
O\,{\sc i}, Si\,{\sc ii}, C\,{\sc ii}, Si\,{\sc iii}, Si\,{\sc iv}, and Fe\,{\sc ii}
together with the 21cm emission profile from the HALOGAS/GBT data in the
range $v_{\rm helio}=+400$ and $+900$ km\,s$^{-1}$. A double-peaked
absorption profile is clearly seen in the lines of O\,{\sc i}
$\lambda 1302.2$, Si\,{\sc ii} $\lambda 1193.3$, and Si\,{\sc iii}
$\lambda 1206.5$. The lines of C\,{\sc ii} $\lambda 1334.5$ and
Si\,{\sc ii} $\lambda 1260.4$ at $v_{\rm helio}\leq +700$ km\,s$^{-1}$
are blended by other intervening absorbers, but the
two absorption components, centered at $+660$ and $+720$ km\,s$^{-1}$,
still are readily visible. A precise modeling of these velocity
components is presented in the next subsection.

We determined total equivalent widths, $W_{\lambda}$, and total logarithmic
column densities, log $N$, by a direct pixel integration over the velocity profiles 
using the apparent optical depth method (Savage \& Sembach 1991).
The values for $W_{\lambda}$ and log $N$ derived in this manner are listed in Table 1,
columns 4 and 5.
In case of non-detections, we determined upper limits for $W_{\lambda}$ and log $N$;
for saturated lines (lines with more than 50 per cent absorption depth)
we interpret the derived values for log $N$ as lower limits (see fifth column
in Table 1).


\begin{figure*}[ht!]
\plotone{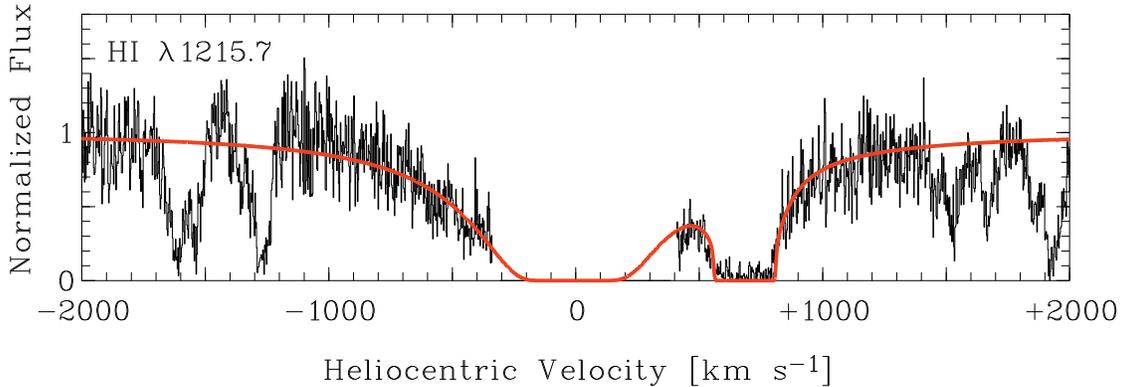}
\caption{
Modeling of the H{\sc i} Ly\,$\alpha$ absorption in the COS spectrum of
2MASS\,J12421031$+$3214268. The red solid line shows the best-fitting
model with Ly\,$\alpha$ absorption from the Milky Way disk and halo gas
near zero velocities and Ly\,$\alpha$ absorption from the NGC\,4631
stream near $+700$ km\,$^{-1}$ (see Sect.\,4.3).
}
\end{figure*}

%

\subsection{Component modeling}

A more advanced approach to derive column densites and gain more
precise information on the velocity structure is the component modeling
method. In this method, we model the shape of the metal absorption by considering 
multiple velocity components and take into account the instrumental line-spread
function of the COS instrument (see Richter et al.\,2013 for a detailed description
of this method). Errors have been
determined by varying $N$ and $b$ in the model in the allowed range, which is constrained
by the residuals between the model spectrum and the data. All modeling results 
are presented in Table 2.

The weak absorption lines from O\,{\sc i} $\lambda 1302.2$ and
Si\,{\sc ii} $\lambda 1304.4$ can be successfully modeled
with two relatively narrow absorption components at $v_1=+655$ km\,s$^{-1}$,
$v_2=+720-730$ km\,s$^{-1}$ and with Doppler parameters $b_1=12-15$ km\,s$^{-1}$,
$b_2=7-10$ km\,s$^{-1}$, hereafter referred to as comp.\,1 and 2. 
The strong line of Si\,{\sc iii} $\lambda 1206.5$ 
exhibits further a broad, absorption wing in the blue, extending
down to $\sim +500$ km\,s$^{-1}$ as well as moderate absorption near 
$\sim +700$ km\,s$^{-1}$, situated just between comp.\,1 and 2.
These features indicate the presence of additional
absorption components for Si\,{\sc iii} (Fig.\,4). 

The observed Si\,{\sc iii} absorption profile can be reproduced in our model
by adding either one, very broad ($b=75$ km\,s$^{-1}$) absorption component
centered at $v=+660$ km\,s$^{-1}$ (Fig.\,4, middle panel) or by adding
two moderately broadened satellite components to comp.\,1 and 2 at
$v_3=+625$ km\,s$^{-1}$ and $v_4=+710$ km\,s$^{-1}$ with
$b_3=b_4=40$ km\,s$^{-1}$ (Fig.\,4, lower panel). 
The blue solid lines shown in the middle 
and lower panels of Fig.\,4, which display the resulting total absorption
profile, do both reproduce the observed absorption profile of 
Si\,{\sc iii} very well. 

The extremely high $b$ value of $75$ km\,s$^{-1}$,
which is required to match the profile with just one additional
component, is unphysically high for a single, photo-ionized circumgalactic 
gas cloud, as neither thermal broadening nor large-scale turbulent gas motions 
are expected to boost the Doppler parameter to such high values (see, e.g.,
Herenz et al.\,2013 for a statistics on $b$ values in Milky-Way CGM clouds).
We therefore reject this model, as it appears unrealistic. The second model with
the two satellite components at $+660$ and $+710$ km\,s$^{-1}$ in Si\,{\sc iii} 
could be naturally explained by a core-envelope structure of the absorbing clouds,
with comp.\,3 and 4 being the outer ionized gas layers (envelopes) of
comp.\,1 and 2. Such multi-phase absorbers are regularly seen
in the local CGM and in circumgalactic Si\,{\sc ii}/Si\,{\sc iii}/Si\,{\sc iv}
systems around other low-redshift galaxies (Richter et al.\,2016,2017; 
Muzahid et al.\,2018).
While we cannot exclude that the velocity structure is even more complex than
what is evident from the COS data, which are limited in spectral resolution and
S/N, we here do not consider any other models with more than four 
absorption components. No valuable information would be gained from
such hypothetical models.
Note that the total Si\,{\sc iii} column density in the weak satellite 
components is constrained to log $N=13.30\pm0.15$ and is independent of 
the actual {\it number} of absorption components. We will further 
discuss the multi-phase nature of Spur\,2 in Sect.\,5.3.

From the modeling (Table 2) follows that the ion column densities
are relatively evenly distributed over the four absorption components.
The total ion column densities (i.e., the sum over all four absorption
components) are given in Table 1, sixth column. They are in good agreement
with the total ion column densities derived from the AOD method
(Table 1, fifth row).

%

\subsection{H\,{\sc i} Ly\,$\alpha$ absorption}

The large velocity separation between the Galactic absorption and absorption
from Spur2 around NGC\,4631 allows us to model the
H\,{\sc i} Ly\,$\alpha$ absorption profile and obtain a precise
estimate of the (pencil-beam) neutral gas column density in Spur 2 from
the UV data.
The H\,{\sc i} Ly\,$\alpha$ absorption profile between
$-2000$ and $+2000$ km\,s$^{-1}$ is shown in Fig.\,5, with
the best-fitting model indicated with the red solid line. For the damped
Milky Way disk absorption trough we assume a single absorption
component centered at zero velocities with log $N$(H\,{\sc i}$)=19.78$,
predominantly constrained by the shape of the extended Lorentzian damping
wing on the blue side of the absorption trough.

For the modeled Ly\,$\alpha$ absorption in Spur 2 in the
range $500-1000$ km\,s$^{-1}$ we assume two neutral absorption components
fixed at the velocities of the two narrow O\,{\sc i} components seen
at $v_1=+655$ km\,s$^{-1}$ and $v_2=+720$ km\,s$^{-1}$ (Table 2).
This approach is justified, because O\,{\sc i} and H\,{\sc i} have
identical ionization potentials and they trace exactly the same gas phase. 
We further assume that the oxygen abundance in both narrow components is 
identical, so that


\begin{equation}
\frac{N_1({\rm OI})}{N_2({\rm OI})}=
\frac{N_1({\rm HI})}{N_2({\rm HI})}.
\end{equation}

%

This approach allows us to fix the column density ratio in the H\,{\sc i}
Ly\,$\alpha$ model to that of O\,{\sc i} and thus only the total H\,{\sc i}
column density remains as free parameter in the model, if a 
constant metallicity is assumed. Doing this,
the total H\,{\sc i} column density in Spur 2 towards 
2MASS\,J12421031$+$3214268 comes out to
log $N$(H\,{\sc i}$)=18.68\pm 0.15$. As can be seen in Fig.\,5,
the resulting model reproduces the observed shape of the observed
Ly\,$\alpha$ absorption very well.

There is no evidence that comp.\,3 and 4 seen in
Si\,{\sc iii} contribute significantly
to the H\,{\sc i} Ly\,$\alpha$ absorption.
This in line with the idea that they
contain predominantly {\it ionized} gas with a
small neutral gas fraction.

%

\begin{deluxetable*}{lcclcclcclccl}
\tabletypesize{\scriptsize}
\tablewidth{0pt}
\tablecaption{Parameters for the absorption-line modeling}
\tablehead{
\colhead{Ion} & 
\colhead{$v_1$} & \colhead{$b_1$} & \colhead{log $N_1$} &
\colhead{$v_2$} & \colhead{$b_2$} & \colhead{log $N_2$} &
\colhead{$v_3$} & \colhead{$b_3$} & \colhead{log $N_3$} &
\colhead{$v_4$} & \colhead{$b_4$} & \colhead{log $N_4$} \\
\colhead{} &
\colhead{[km\,s$^{-1}$]} & \colhead{[km\,s$^{-1}$]} & \colhead{} &
\colhead{[km\,s$^{-1}$]} & \colhead{[km\,s$^{-1}$]} & \colhead{} &
\colhead{[km\,s$^{-1}$]} & \colhead{[km\,s$^{-1}$]} & \colhead{} &
\colhead{[km\,s$^{-1}$]} & \colhead{[km\,s$^{-1}$]} & \colhead{} 
}
\startdata
O\,{\sc i}    & +665 & 12 & 14.35 &   +720 &  7 & 14.10 & +625 & ... & ...          & +710 & ... & ...         \\
C\,{\sc ii}   & +655 & 15 & 14.70\tablenotemark{a}
                                  &   +730 & 10 & 14.20 & +625 &  40 & 14.00\tablenotemark{a}
                                                                                    & +710 & 40 & 13.80        \\  
Si\,{\sc ii}  & +655 & 15 & 13.60 &   +730 & 10 & 13.10 & +625 &  40 & $\leq 13.40$ & +710 & 40 & $\leq 13.10$ \\ 
Si\,{\sc iii} & +655 & 15 & 13.35 &   +730 & 10 & 13.30 & +625 &  40 & 13.00        & +710 & 40 & 13.00        \\
H\,{\sc i}    & +655 & 25 & 18.50 &   +730 & 20 & 18.25 & +625 & ... & ...          & +710 & ... & ...         
\enddata
\tablenotetext{a}{Comp.\,1 blended by intervening H\,{\sc i} Ly\,$\zeta$ absorption at $z=0.043677$.}
\end{deluxetable*}


\subsection{21cm emission}

Fig.\,1 indicates that the QSO sightline passes the tidal streams around NGC\,4631 
just beyond the outer boundaries of the 21cm intensity contours of Spur 2. 
The H\,{\sc i} column density
measured from the Ly\,$\alpha$ absorption (log $N$(H\,{\sc i}$)=18.68$; see
above) is below the detection limit of the high-resolution HALOGAS/WSRT data,
suggesting that the 2MASS\,J12421031$+$3214268 sightline traces the 
diffuse outer envelope of Spur 2.
In the lowest panel of Fig.\,3 we show the 21cm emission spectrum
in the range $v_{\rm helio}=400-900$ km\,$^{-1}$ from
the HALOGAS/WSRT data cube centered on the position of 2MASS\,J12421031$+$3214268
(blue solid line). No significant emission is seen in the WSRT data (as expected),
but we can place an upper limit on the H\,{\sc i} column density of 
log $N$(H\,{\sc i}$)=19.15$. This is in line with the measured 
value of log $N$(H\,{\sc i}$)=18.68$ from the
Ly\,$\alpha$ absorption. With the red solid line we have overlaid the 
21cm emission spectrum from the {\it combined} WSRT+GBT data, which has a
substantially larger effective beam size at the position of the QSO 
(see discussion in Sect.\,3.2). A weak ($T_{\rm B,max}\approx 0.25$ K)  
emission peak is seen at $+680$ km\,$^{-1}$, corresponding to 
log $N$(H\,{\sc i}$)=19.65$. This feature is a result of
beam-smearing effects  near the 21cm boundary of Spur 2, caused 
by the low angular resolution of the combined WSRT+GBT data. 
A substantial fraction of
the combined WSRT+GBT beam centered on 2MASS$-$J12421031$+$3214268 is filled 
with 21cm emission from the inner region of Spur 2, leading to the 
observed emission feature.

%

\section{Discussion}

\subsection{Metallicity of the gas}

Because of the identical ionization potentials of neutral
oxygen and hydrogen and the unimportance of dust depletion
effects for O, the O\,{\sc i}/H\,{\sc i} ratio is a reliable
indicator for the overall $\alpha$ abundance in neutral gas.
For log $N$(H\,{\sc i}$)>18$ only very small ionization corrections 
($<0.1$ dex) apply (see, e.g., Richter et al.\,2018).

From the O\,{\sc i} and H\,{\sc i} column densities listed
in Table 2 and and the ionization model 
described in Sect.\,5.3 we determine a metallicity of the gas of
$0.13^{+0.07}_{-0.05}$ solar or [$\alpha$/H$]=-0.90\pm 0.16$,
where we assume a solar oxygen reference abundance 
of (O/H)$_{\sun}=-3.31 \pm 0.05$ (Asplund et al.\,2009). 
Such a metallicity is at the low end
of the $\alpha$-abundance distribution observed in massive spiral
galaxies (Pilyugin et al.\,2014), but is very similar to the values found in the main
body of the Magellanic Stream (MS) and other HVCs in the halo
of the Milky Way (Fox et al.\,2013; Richter et al.\,2001; Tripp et al.\,2003).

Interestingly, the derived metallicity of $[\alpha$/H$]=-0.90\pm 0.16$
in Spur 2 falls right into the metallicity gap seen in LLS at low redshifts 
(Lehner et al.\,2013; their Fig.\,3). In terms of metallicity, Spur 2 around NGC\,4631
and the MW HVCs thus appear to be atpypical for optically thick H\,{\sc i} 
absorbers in the local Universe. It remains to be explored in future studies 
whether this is due to the limited statistics on the metallicity of low-redshift LLS or 
could be related to the tidal origin of Spur 2, the MS and and other MW HVCs.

\subsection{Possible sources of Spur 2}

In the following, we discuss the various possible sources of 
Spur 2 by combining our metallicity measurement with other available
information.

\subsubsection{The outer disk of NGC\,4631}

One possible scenario is that Spur 2 represents gas that has been 
tidally torn out of the outer, metal-poor gas disk of NGC\,4631 during a 
close encounter of NGC\,4627 or another satellite galaxy.
Spur 2 does appear to connect spatially and kinematically with the outer 
NGC\,4631 H\,{\sc i} disk over a large section of the disk's major axis 
(Rand 1994), which supports a NGC\,4631 disk origin.

The derived metallicity of [$\alpha$/H$]=-0.90\pm 0.16$ in Spur 2 is, however,
significantly below the interstellar oxygen abundance in the 
central region of the NGC\,4631 disk ([M/H$]=-0.30\pm 0.06$; Pilyugin et al.\,2014).
If we apply the observed radial metallicity gradient for NGC\,4631
of $-0.0194\pm 0.0073$ dex/kpc (Pilyugin et al.\,2014), the derived
$\alpha$ abundance in Spur 2 would match the metallicity of NGC\,4631
only in the very outer disk at $r>20$ kpc. Most likely, the 
extrapolation to such large radii is not valid, as the metallicity
gradients in disk galaxies appear to flatten out at large radii (Werk et al.\,2011).
Therefore, the measured low $\alpha$ abundance in Spur 2 does not favor an
origin in the disk of NGC\,4631.

We further note that the initial metallicity of Spur 2, at the time when the gas 
was separated from its source galaxy, could have been even lower than what 
is observed today. This is because the tidal features potentially have started 
to mix with the ambient hot coronal gas around NGC\,4631, which is expected
to be fed by outflowing, chemically enriched material from star-forming regions 
and thus should be more metal-rich (see review by Fraternali 2017).

\subsubsection{NGC\,4627 or other satellite galaxies}

Alternatively, Spur 2 could represents material that
has been stripped from a gas-rich companion of NGC\,4631.
The stripped gas would then be located
in the halo of NGC\,4631 and currently being
accreted onto the disk, which would explain its spatial and kinematical connection
to the H\,{\sc i} disk.
Possible galaxy candidates are the dE satellite
NGC\,4627, or the tidally disrupted satellite
galaxy that forms the recently detected stellar stream
around NGC\,4631 (Mart\'inez-Delgado et al.\,2015),
or another recently disrupted gas-rich satellite galaxy. 

Combes (1978) has presented a detailed model of the tidal
interactions between NGC\,4631, NGC\,4656, and NGC\,4627 based
on the early 21cm observations presented in Weliachev et al.\,(1978).
From her model, which assumes a parabolic encounter of NGC\,4656 with NGC\,4631,
follows that basically all gaseous material that is seen in Spurs 1, 4, and 5
and in the gas near NGC\,4656 stems from the NGC\,4631 disk. 
Spurs 2 and 3, however, cannot be explained with this model. 
Combes proposes that the gas-rich progenitor
galaxy of NGC\,4627 has lost basically all its interstellar neutral gas
in a close encounter with NGC\,4631, material that is now 
seen in 21cm emission as Spurs 2 and 3.
While the Combes model disfavours the nearby companion NGC\,4656 as
source for Spur 2, it is worth noticing that the measured
$\alpha$ abundance in Spur 2 coincides well with the interstellar
metallicity in the inner 10 kpc of NGC\,4656 (Pilyugin et al.\,2014).
If some of the neutral material in the gaseous bridge between 
NGC\,4631 and NGC\,4656 is connected to the extended H\,{\sc i}
halo of NGC\,4656, then it cannot be excluded that the absorption
towards MASS\,J12421031$+$3214268 is related to gas that originally 
stems from NGC\,4656.
Clearly, it would be highly desirable to re-model the tidal interactions
between NGC\,4631, NGC\,4656, and NGC\,4627 using state-of-the art
simulation codes together with the most recent observational constraints as input
parameters to further clarify these issues.


\begin{deluxetable*}{lccclcccr}
\tabletypesize{\small}
\tablewidth{0pt}
\tablecaption{Results from ionization models}
\tablehead{
\colhead{Comp.} & \colhead{$v$} & \colhead{log $N$(H\,{\sc i})} & \colhead{log $\delta$(Si)} & 
\colhead{UV model} & \colhead{log $N$(H\,{\sc ii})} &
\colhead{log $n_{\rm H}$} & \colhead{log $f_{\rm HI}$} & \colhead{$l$} \\
\colhead{} & \colhead{[km\,s$^{-1}$]} & \colhead{} & \colhead{[dex]} &
\colhead{} & \colhead{} & \colhead{} & \colhead{} & \colhead{[pc]} 
}
\startdata
1 &   $+655$ & $18.45$  & $-0.22$ & UVB     & $19.25$ & $-2.18$ & $-0.86$ &  $1000$ \\
  &          &          &         & UVB+G50 & $19.25$ & $-1.40$ & $-0.86$ &   $166$ \\
  &          &          &         & UVB+G10 & $19.25$ & $-0.64$ & $-0.86$ &    $29$ \\
2 &   $+730$ & $18.30$  & $-0.22$ & UVB     & $18.55$ & $-0.87$ & $-0.53$ &    $10$ \\
  &          &          &         & UVB+G50 & $18.55$ & $-0.09$ & $-0.53$ &     $2$ \\
  &          &          &         & UVB+G10 & $18.55$ & $+0.67$ & $-0.53$ &   $0.3$ \\
\enddata
\tablenotetext{a}{Only the gas phase traced by O\,{\sc i}, C\,{\sc ii}, and Si\,{\sc ii} is considered.}
\tablenotetext{b}{Si depletion value.}
\tablenotetext{c}{UV flux models considered: 1) EG=extragalactic UV background at z=0; 
2) EG+G50kpc=extragalactic UV background + galaxy contribution assuming $d=50$ kpc 
impact parameter; 3) EG+G10kpc=EG+galaxy at $d=10$ kpc.}
\end{deluxetable*}


It is an interesting fact, that the derived oxygen abundance of [M/H$]=-0.90$
in Spur 2 matches almost exactly the metallicity derived for
the stellar stream ([M/H$]=-0.92$; Tanaka et al.\,2017).
The uncertainty for the stellar abundance is quite large, however, with
a 90 percent confidence interval between $-1.46$ and $-0.51$
(see Tanaka et al.\,2017). Therefore, this apparent similarity does not provide
strong constraints on a possible common origin.
There are, in fact, several arguments that speak against a connection
between the stellar and the gaseous streams. From the tidal model
of Mart\'inez-Delgado et al.\,(2015) it follows, for instance,
that the stellar stream must be old, with a preferred
age of $\sim 3.5$ Gyr to qualitatively match the observed 
morphology of the stream. Spur 2 must be substantially younger, however,
because the stream's gas is expected to interact with the ambient hot 
coronal gas (e.g., Heitsch \& Putman 2009). As a result, the gas 
should either be disrupted and incorporated into the corona, or being
accreted onto the disk of NGC\,4631 on times scales much smaller 
than $3.5$ Gyr (typical disruption time-scales
would be a few 100 Myr; Heitsch \& Putman 2009), i.e., the gas structure 
would not survive for such a long time in such a massive galaxy halo.
Another result of the tidal model from Mart\'inez-Delgado et al.\,(2015) 
is that the total inital stellar mass
of the accreted dwarf galaxy must have been small, $< 6\times 10^8 M_{\sun}$.
This value is, however, smaller than the neutral gas mass of Spur 2 alone
($\sim 8\times 10^8) M_{\sun}$ (Rand 1994).
In conclusion, Spur 2 is almost certainly not related to the stellar stream.

%

\begin{figure}[t!]
\epsscale{1.1}
\plotone{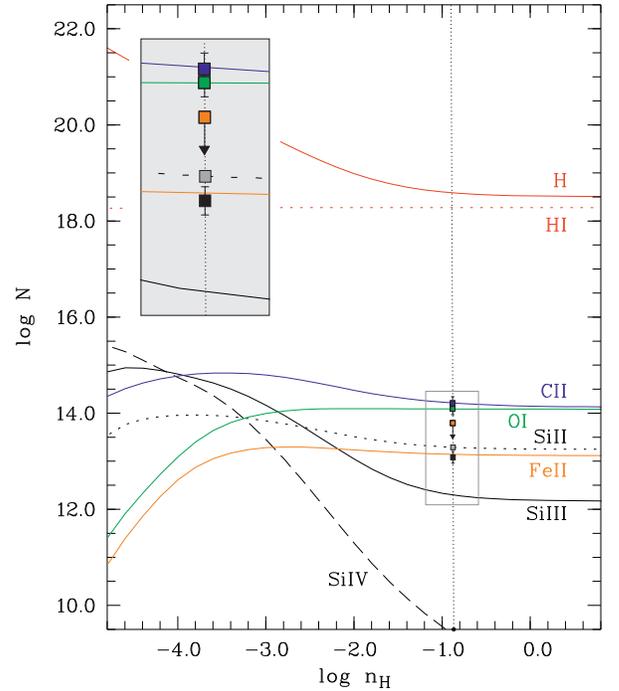}
\caption{
Predicted column densities (solid lines) for various ions as a function
of the gas density from the Cloudy ionization model for comp.\,2, assuming
just the UV background as ionization source (see Sect.\,5.3; Table 3).
The observed values (filled boxes) fit best to a gas density of
log $n_{\rm H}=-0.87$. The gray box symbol indicates the Si\,{\sc ii}
column density assuming a depletion of log $\delta$(Si$)=-0.22$ dex.
The gray-shaded inset displays a zoom-in into the region indicated
with the gray solid frame.
}
\end{figure}


\subsection{Ionization conditions and total gas mass}

We now discuss the ionization conditions in the different
gas components in more detail by analyzing the observed
O\,{\sc i}/C\,{\sc ii}/Si\,{\sc ii} column-density
ratios with a grid of Cloudy ionization models (Ferland et al.\,2013)
that are based on the H\,{\sc i} column densities (limits) derived for
comp.\,1 and 2 (Table 2).

We follow the strategy outlined in our previous
studies (e.g., Richter et al.\,2009, 2013, 2016)
and model the ioniziation conditions in Spur 2 assuming the
absorbers to be a simple gas slab that is illuminated from 
both sides. In principle, the strength and shape of the illuminating radiation
field in a CGM cloud of a star-forming galaxy depends on the
overall extragalactic UV background field at that redshift and the 
location of the cloud with respect to the UV-emitting 
star-forming disk. Since we are lacking any accurate information
on the distance of Spur 2 to the NGC\,4631 stellar disk and the shape/intensity
of the local radiation field, we consider three models
with different assumptions for the local radiation field in Spur 2.
In the first model, we use only the extragalactic UV background (UVB)
field at $z=0$ normalized to a photoionization rate of 
log $\Gamma=-13.34$ (see discussion in Kollmeier et al.\,2014; Wakker et al\,2015; 
Shull et al.\,2015). The second model assumes a standard UVB field together 
with an additional Milky-Way type radiation field (Fox et al.\,2005, 2014) 
at a galactocentric distance of $d=50$ kpc (UVB+G50), as explained 
in Fox et al.\,(2014) and Richter et al.\,(2009). 
The third model assumes a UVB field with an additional Milky-Way type radiation 
field at a galactocentric distance of $d=10$ kpc (UVB+G10). 

For the H\,{\sc i} column densities
in components 1 and 2 the expected ionization correction to
determine (O/H) from the measured O\,{\sc i}/H\,{\sc i} is only
$0.03$ dex. For the derived $\alpha$ abundance in the gas 
([$\alpha$/H$]=-0.90$), Cloudy then delivers for each of the
three assumed radiation fields (UVB, UVB+G50, UVB+G10)
predictions for the column densities of the other ions as
a function of the local gas density, log $n_{\rm H}$ (see Fig.\,6
for a Cloudy-example plot).
By matching the observed column densities
with the Cloudy-model predictions for UVB, UVB+G50, UVB+G10 
we obtain for each component estimates for the ionized gas column density,
$N$(H\,{\sc ii}), the gas density, the neutral gas fraction, 
$f_{\rm HI}=N$(H\,{\sc i}$)/[N$(H\,{\sc i}$)+N$(H\,{\sc ii})],
and the thickness of the absorbing gas layer, $l=N$(H$)/n_{\rm H}$. 
The modeling results are listed in Table 3.

Note that we do not include Si\,{\sc iii} in our modeling for comp.\,1 and 2, 
as Si\,{\sc iii} and O\,{\sc i}/Si\,{\sc ii} are not expected to arise
in the same gas phase (see discussions in Muzahid et al.\,2018; Richter et al.\,2016). 
The fine-tuning of the Si\,{\sc ii}/O\,{\sc i} ratio in comp.\,1 and 2
in the Cloudy model was used to determine the dust depletion of Si, as will
be described in Sect.\,5.4.

From the Cloudy modeling follows that the density in comp.\, 1 lies 
between log $n_{\rm H}=-2.18$ and $-0.64$, which indicates an absorber
size between $\sim 30$ pc and $1$ kpc, depending on 
the assumed radiation field (see Table 3). The gas density
in comp.\,2 is predicted to be one order of magntiude higher
(log $n_{\rm H}=-0.87$ to $+0.67$), indicating that comp.\,2
repesents a small gas clump with a size $\leq 10$ pc.
It is important to keep in mind that these values are based on the 
above outlined (idealized) assumptions on the radiation field and absorber
geometry. Consequently, they should be interpreted with some caution.
If Spur 2 has a similarly large distance to the disk of NGC\,4631 as 
the main body of the MS to the MW disk ($d=50-100$ kpc), then the 
UVB+G50 model applies, indicating that the absorbing clouds have pc-scale
sizes at moderate densities. In this model, which we regard as the
most realistic one given the geometrical configuration of the Spurs around
NGC\,4631, the derived gas densities and clump sizes are very similar
to those seen in the MS and other HVCs in the Milky Way halo
(Richter et al.\,2009, 2013, 2018), pointing at
comparable physical conditions.

While we cannot provide any Cloudy modeling for the satellite
components 3 and 4, we can give a lower limit for the total
hydrogen column density, $N$(H$)=N$(H\,{\sc i}$)+N$(H\,{\sc ii}),
in each of these components. For this, we transform the observed
Si\,{\sc iii} column density in comp.\,3 and 4 (Table 2) 
into $N$(H) assuming that the silicon abundance in the gas is $0.13$ solar, 
similar to what is measured in comp.\,1 and 2. With the 
resulting value log (Si/H$)=-5.29$ in Spur 2 we find that
log $N$(H$)\geq 18.29$ in each of the two comp.\,3 and 4,
or log $N$(H$)\geq 18.59$ in the two components together.

Combining the results for all four components, we derive for 
the mean neutral gas fraction in Spur 2 along the
2MASS\,J12421031$+$3214268 sightline a value of
$\langle f_{\rm HI} \rangle =0.14$, indicating that there
is $\sim 7$ times more ionized hydrogen than neutral hydrogen
in the stream in this direction. This ratio may not be typical
for the entire tidal stream. This is because the 2MASS\,J12421031$+$3214268
sightline passes the stream just at the boundaries of the 21cm contours
where the neutral gas column is subtantially smaller than in the inner regions
of Spur 2 (see Fig.\,1, upper panel). 
Yet, the Cloudy results imply that Spur 2 contains substantial
amounts of ionized gas that most likely brings its total gas mass
to a few times $10^{9} M_{\sun}$. This value is 
comparable to the total gas mass estimated for the Magellanic Stream in
the Milky Way halo (Fox et al.\,2014).

\subsection{Dust content}

As has been demonstrated in the case of the Magellanic Stream, interstellar dust grains
can survive the tidal stripping proccess in merging galaxies (Lu et al.\,1998;
Gibson et al.\,2001; Fox et al.\,2013; Richter et al.\,2013, 2018).
To investigate the dust abundance in Spur 2, we compare in detail
the abundance of the two $\alpha$ elements Si and O in our Cloudy model
for components 1 and 2. The Si/O abundance ratio is a useful dust indicator.
This is because even in relatively diffuse gas in the Milky Way's ISM and CGM
Si is moderately depleted into dust grains, while O is not (e.g., Savage \& Sembach 1996).

The Cloudy models for comp.\, 1 and 2 predict Si\,{\sc ii} column densities
that are systematically
higher than those observed, suggesting that Si indeed is depleted into dust grains.
The effect is small, however, with an absolute depletion value of only 
$|$log $\delta$(Si$)|=0.22$ dex (Table 3, third row).
This value corresponds roughly to the systematic errors in the Cloudy model,
so in the following we regard $|$log $\delta$(Si$)|\leq 0.2$ dex as a realistic upper
limit for the Si depletion in Spur2.

Interestingly, this upper limit is still substantially smaller than than the
characteristic Si depletion values recently
observed in the Magellanic Stream, its Leading Arm, and the Magellanic Bridge
($|$log $\delta$(Si$)|\geq 0.5$ dex; Fox et al.\,2013; 
Richter et al.\,2013, 2018; Lehner 2002).
As discussed in Richter et al.\,(2018), the level of depletion in
star-less tidal streams possibly reflect the initial conditions in the
interstellar medium of the stream's source galaxy, if the gas has been
removed ``quiescently'' from its host (e.g., by gravitational forces), but not
by dust-destroying energetic events (e.g., by stellar winds or supernova-driven outflows).
Therefore, and independently of whether Spur 2 originally stems from a 
satellite galaxy or from the outer disk of NGC\,4631, the 
small absolute Si depletion value in Spur 2 indicates an overall low dust content 
in the original host environment of the gas.

%

\section{Summary and conclusions}

We have analyzed HST/COS UV spectral data of the background QSO 2MASS\,J12421031$+$3214268
and 21cm observations from the HALOGAS project with support from new GBT observations
to study the chemical composition and physical conditions in one of the tidal
gas streams (Spur 2) around the Whale galaxy NGC\,4631. The main results of
our study can be summarized as follows:\\
\\
1.) Strong metal- and hydrogen absorption related to Spur 2 
    is detected in the COS spectrum of 2MASS\,J12421031$+$3214268
    in the velocity range between $+550$ and $800$ km\,s$^{-1}$, although
    the sightlines passes the stream just beyond the outer 21cm boundaries.
    Detected atoms/ions include H\,{\sc i}, O\,{\sc i}, C\,{\sc ii}, Si\,{\sc ii},
    and Si\,{\sc iii}.\\
\\
2.) Four individual velocity components are identified in the metal absorption with
    two narrow components ($b=7-15$ km\,s$^{-1}$) and two broader components
    ($b=40$ km\,s$^{-1}$). The broader components, which are seen only in Si\,{\sc iii}, 
    possibly trace more extended, ionized gas layers that surround the denser gas structures
    sampled by the low-ionzation states.  
    This complex absorption pattern suggests a multi-phase nature of the tidal gas stream,
    similar to what is seen in the Magellanic Stream in the outer Milky Way halo.\\
\\
3.) A fit to the H\,{\sc i} Ly\,$\alpha$ absorption in the 2MASS\,J12421031$+$3214268
    spectrum yields a neutral gas column density of log $N$(H\,{\sc i}$)=18.68\pm0.15$ in
    Spur 2, which is below the detection limit in the 21cm data. 
    From a set of Cloudy ionization models it follows that the total hydrogen column 
    in Spur 2 towards 2MASS\,J12421031$+$3214268 is log $N$(H$)=19.46$, indicating
    a sightline-averaged neutral gas fraction of $14$ percent. 
    The gas density in the narrow absorption components range between 
    log $n_{\rm H}=-2.18$ to $+0.67$, suggesting that these absorbers are CGM cloudlets
    on sub-kpc scales.\\
\\
4.) From the unsaturated O\,{\sc i} absorption, together with the derived H\,{\sc i}
    column density and the Cloudy ionization correction, we derive an $\alpha$ abundance
    in the gas of ($\alpha$/H$)=0.13^{+0.05}_{-0.04}$ solar ([$\alpha$/H$]=-0.90\pm 0.16$).
    This value is lower than the abundance in the NGC\,4631 gas disk, even if the radial
    abundance gradient is taken into account. Although the kinematic connection of Spur 2
    with the gas disk (Rand 1994) points toward a NGC\,4631 disk origin, the low metallicity
    in the gas rather favors a satellite-galaxy origin for Spur 2.
    From the Cloudy models we derive an upper limit for the Si depletion in Spur 2 of
    $|$log $\delta$(Si$)|\leq 0.2$, suggesting that the gas has a very low dust abundance.\\
\\
We conclude that the observed properties of Spur 2 favor the scenario,
in which metal- and dust-poor gas has been tidally torn out of 
a (formerly) gas-rich satellite galaxy in a recent encounter.
The dE satellite NGC\,4627 at 6 kpc distance from NGC\,4631 thus represents 
a particularly promising candidate for the encountering galaxy.

The ionization conditions in the gas observed towards 2MASS\,J12421031$+$3214268
further indicate that Spur 2 contains $\sim 7$  times more ionized than neutral gas, 
which lifts its total mass to a value of a few times $10^{9} M_{\sun}$.
If similar ionization fractions apply to the other spurs around NGC\,4631, however, then
the combined total gas mass of the NGC\,4631 tidal gas streams may easily exceed
$10^{10}$ $M_{\sun}$.
This extremely large reservoir of relatively cool ($T<T_{\rm vir}$), metal-poor
circumgalactic gas will potentially be accreted by NGC\,4631 over the next few hundred Myr
and will boost its star-formation rate.
Additional information on the chemical composition and physical conditions of the
other four spurs of the Whale galaxy's tidal stream system
would require a larger number of UV-bright background AGN distributed around NGC\,4631,
which unfortunately are not available. From what we know so far, however,
the NGC\,4631 gas spurs reflect many of the properties seen in the 
the Magellanic Stream and its various components (Mathewson et al.\,1974; 
Putman et al.\,1998; Br\"uns et al.\,2005; D'Onghia \& Fox 2016).

While the analysis of individual QSO sightlines around nearby galaxy mergers provide
interesting information on these particular systems, a more systematic approach
in studying the neutral and ionized gas distribution in group environments is
highly desirable. As demonstrated here, absorption spectroscopy in the UV represents a
powerful method to characterize the multi-phase nature of tidal gas streams,
to estimate their chemical composition, and their ionized gas mass along individual sightlines.
The combination of a larger sample of QSO sightlines passing nearby galaxy groups, data from future
deep 21cm surveys, deep X-ray observations, and high-resolution hydrodynamical simulations
holds the prospect of substantially improving our understanding of the overall importance
of massive tidal gas streams in the context of galaxy formation.

\acknowledgments

Based on observations obtained under program 14085 with the NASA/ESA
Hubble Space Telescope, which is operated by the Space
Telescope Science Institute (STScI) for the Association of
Universities for Research in Astronomy, Inc., under NASA
contract NAS5D26555.

%

\section*{REFERENCES}
\begin{footnotesize}

\noindent
\\
Ann, H.B., Seo, M.S., \& Baek, S.-J. 2011, 
Journal of Korean Astronomical Society, 44, 23
\noindent
\\
Asplund, M., Grevesse, N., Jacques Sauval, A., \& Scott, P. 2009, ARA\&A, 47, 481
\noindent
\\
Barger, K.A., Haffner, L.M., \& Bland-Hawthorn, J. 2013 ApJ, 771, 132
\noindent
\\
Bendo, G. J., Dale, D. A., Draine, B. T., et al. 2006, ApJ, 652, 283
\noindent
\\
Besla, G., Kallivayalil, N., Hernquist, L., et al. 2010, ApJ, 721, L97
\noindent
\\
Besla, G., Kallivayalil, N., Hernquist, L., et al. 2012, MNRAS, 421, 2109
\noindent
\\
Br\"uns, C, Kerp, J., Staveley Smith, L., et al. 2005, A\&A, 432, 45
\noindent
\\
Chynoweth, K. M., Langston, G. I., Yun, M. S., et al. 2008, AJ, 135, 1983
\noindent
\\
Combes, F. 1978, A\&A, 65, 47
\noindent
\\
Connors, T.W., Kawata, D., \& Gibson, B.K. 2006, MNRAS, 371, 108
\noindent
\\
Dave\'e, R., Finlator, K., \& Oppenheimer, B.D. 2012, MNRAS, 421, 98
\noindent
\\
Debes, J.H., Becker, G., Roman-Duval, J., et al. 2016, Instrument Science Report COS, 15
\noindent
\\
Diaz, J.D., \& Bekki, K. 2011, MNRAS, 413, 2015
\noindent
\\
Diaz, J.D., \& Bekki, K. 2012, ApJ, 750, 36
\noindent
\\
Di\,Teodoro, E.M. \& Fraternali, F. 2014, A\&A, 567, 68
\noindent
\\
D'Onghia, E. \& Fox, A.J. 2016, ARA\&A, 54, 363
\noindent
\\
Fabbiano, G. \& Trinchieri, G. 1987, ApJ, 315, 46
\noindent
\\
Ferland, G. J., Porter, R. L., van Hoof, P. A. M., et al. 2013, RMxAA, 49, 137
\noindent
\\
Finlator, K. 2017, ASSL, 430, 221
\noindent
\\
Fontana, A., \& Ballester, P. 1995, ESO Messenger, 80, 37
\noindent
\\
Fox, A.J., Wakker, B.P.; Savage, B.D., et al. 2005, ApJ, 630, 332
\noindent
\\
Fox, A.J., Richter, P., Wakker, B.P., et al. 2013, ApJ, 772, 110
\noindent
\\
Fox, A.J., Wakker, B.P., Barger, A., et al. 2014, ApJ, 787, 147
\noindent
\\
Fraternali, F. \& Binney, J.J. 2008, MNRAS, 386, 935
\noindent
\\
Fraternali, F. 2017, ASSL, 430, 323
\noindent
\\
Gardiner, L.T. \& Noguchi, M. 1996, MNRAS, 278, 191
\noindent
\\
Gentile, G., J\'ozsa, G. I. G., Serra, P., et al. 2013, A\&A, 554, A125
\noindent
\\
Gibson, B.K., Giroux, M.L., Penton, S.V., Stocke, J.T., Shull, J.M., 
\& Tumlinson, J. 2001, AJ, 122, 3280
\\
Giuricin, G., Marinoni, C., Ceriani, L., \& Pisani, A. 2000, ApJ, 543, 178
\noindent
\\
Green, J. C., Froning, C. S., Osterman, S., et al. 2012, ApJ, 744, 60
\noindent
\\
Heald, G., J\'ozsa, G., Serra, P., et al. 2011, A\&A, 526, A118
\noindent
\\
Heitsch, F. \& Putman, M. E. 2009, ApJ, 698, 1485
\noindent
\\
Herenz, P., Richter, P., Charlton, J.C., \& Masiero, J.R. 2013, A\&A, 550, A87
\noindent
\\
Kamphuis, P., Rand, R. J., J\'ozsa, G. I. G., et al. 2013, MNRAS, 434, 2069
\noindent
\\
Kacprzak, G. 2017, ASSL, 430, 145
\noindent
\\
Kennicutt, R.C., Lee, J.C., Funes S.J., J.G., Sakai, S., 
\& Akiyama, S. 2008, ApJS, 178, 247
\noindent
\\
Kollmeier, J.A., Weinberg, D.H., Oppenheimer, B.D., et al.\,2014, ApJ, 789, L32
\noindent
\\
Lehner, N. 2002, ApJ, 578, 126
\noindent
\\
Lehner, N., Howk, J.C., Tripp, T.M., et al.\,2013, ApJ, 770, 138
\noindent
\\
Lu, L., Sargent, W.L.W., Savage, B.D., Wakker, B.P., Sembach, K.R., 
\& Oosterloo, T.A. 1998, AJ, 115, 162
\noindent
\\
Martin, C. \& Kern, B. 2001, ApJ, 555, 258
\noindent
\\
Mart\'inez-Delgado, D., D'Onghia, E., Chonis, T.S., et al. 2015, AJ, 150, 116
\noindent
\\
Mathewson, D.S., Cleary, M.N., \& Murray, J.D. 1974, ApJ, 190, 291
\noindent
\\
Melo, V. \& Mu\~noz-Tu\~n\'on, C. 2002, in Astronomical Society of 
the Paciﬁc Conference Series, Vol. 282, Galaxies: the Third Dimension, 
ed. M. Rosada, L. Binette, \& L. Arias, 338
\noindent
\\
Morton, D. C. 2003, ApJS, 149, 205
\noindent
\\
Muzahid, S., Fonseca, G., Roberts, A., et al.\,2018, MNRAS, 476, 4965
\noindent
\\
Neininger, N. \& Dumke, M. 1999, Proceedings of the National Academy 
of Science, 96, 5360
\noindent
\\
Nidever, D. L., Majewski, S. R., Burton, W. B., \& Nigra, L. 2010,
ApJ, 723, 1618
\noindent
\\
Pilyugin, L.S., Grebel, E.K., \& Kniazev, A.Y. 2014, AJ, 147, 131
\noindent
\\
Pisano, D.J. 2014, AJ, 147, 48
\noindent
\\
Radburn-Smith, D.J., de Jong, R.S., Seth, A.C., et al. 2011, ApJS, 195, 18
\noindent
\\
Rand, R. 1994, A\&A, 285, 833
\noindent
\\
Richter, P., Savage, B.D., Wakker, B.P., Sembach, K.R., \& Kalberla, P.M.W. 2001, ApJ, 549, 281
\noindent
\\
Richter, P., Charlton, J. C., Fangano, A. P. M., Ben Bekhti, N.,
\& Masiero, J. R. 2009, ApJ, 695, 1631
\noindent
\\
Richter, P., Fox, A. J., Wakker, B. P., et al. 2013, ApJ, 772, 111
\noindent
\\
Richter, P., Wakker, B. P., Fechener, C., Herenz, P., Tepper-Garc\'ia, T., 
\& Fox, A. J. 2016, A\&A, 590, A68
\noindent
\\
Richter, P., Nuza, S. E., Fox, A. J., et al.\,2017, A\&A, 607, A48
\noindent
\\
Richter, P. 2017, ASSL, 430, 15
\noindent
\\
Richter, P., Fox, A. J., Wakker, B. P., et al. 2018, ApJ, 865, 145
\noindent
\\
Roberts, M. S. 1968, ApJ, 151, 117
\noindent
\\
S\'anchezr-Almeida, J. 2017, ASSL, 430, 67
\noindent
\\
Salem, M., Besla, G., Bryan, G., et al.\,2015, ApJ, 815, 77
\noindent
\\
Sanders, D.B., Mazzarella, J.M., Kim, D.-C., Surace, J.A. 
\& Soifer, B.T. 2003, AJ, 126, 1607
\noindent
\\
Savage, B. D., \& Sembach, K. R. 1991, ApJ, 379, 245
\noindent
\\
Savage, B.D. \& Sembach, K.R. 1996, ARA\&A, 34, 279
\noindent
\\
Shull, J.M., Moloney, J., Danforth, C.W., \& Tilton, E.M. 2015, ApJ, 811, 3
\noindent
\\
Tanaka, M., Chiba, M., \& Komiyama, Y. 2017, ApJ, 842, 127
\noindent
\\
Tripp, T.M., Wakker, B.P., Jenkins, E.B., et al. 2003, AJ, 125, 3122
\noindent
\\
Wang, Q.D., Walterbos, R.A.M., Steakley, M.F., Norman, C.A., \& 
Braun, R. 1995, ApJ, 439, 176
\noindent
\\
Wakker, B.P., Hernandez, A.K., French, D., et al. 2015, ApJ, 840, 14
\noindent
\\
Weliachew, L., Sancisi, R., \& Gu\'elin, M. 1978n A\&A, 65, 37
\noindent
\\
Werk, J.K., Putman, M.E., Meurer, G.R., \& Santiago-Figueroa, 
N. 2011, ApJ, 735, 71
\noindent
\\
Yamasaki, N.Y., Sato, K., Mitsuishi, I., \& Ohashi, T. 2009, PASJ, 61, 291
\noindent
\\
Yun, M.S., Ho, P.T.P., \& Lo, K.Y. 1994, Nature, 372, 530
\noindent
\\
Zschaechner, L.K., Rand, R.J., Heald, G.H., Gentile, G., \& 
Kamphuis, P. 2011, ApJ, 740, 35

\end{footnotesize}

%

\newpage

\appendix

\section{Combination of WSRT and GBT data}

The CASA task `feather', which was used to combine the GBT with the WSRT data, 
first regrids the GBT data to match the spectral and 
spatial resolution of the WSRT data before Fourier transforming each image to a 
gridded $\it{u-v}$ plane. The $\it{u-v}$ data of the GBT are then scaled by the ratio of 
the volume of the two beams

\begin{equation}\label{eq:beamFrac}
\alpha = \frac{\Omega_{WSRT}}{\Omega_{GBT}},
\end{equation}

where we find $\alpha$ to equal $0.0059$. This factor accounts for the difference in 
flux of the two data sets based solely on the differences in resolution. 
The WSRT $\it{u-v}$ data are then scaled by the factor

\begin{equation}\label{eq:weight}
\beta = 1-\mathfrak{F}[B_{GBT}],
\end{equation}

where the second term represents the Fourier transform of the GBT beam as a function 
of sky angles. The two scaled $\it{u-v}$ data sets are then summed and Fourier transformed 
back to the image plane. The scaling depicted in equation (A2) ensures the 
effects of the poorly sampled low spatial frequencies in the WSRT data are smoothly 
removed before the well-sampled low spatial frequencies provided by the GBT data are added.

Before computing the zeroth (top) and first (bottom) statistical moments of the cube as shown 
in Figure 1, the combined cube was masked at the 3$\sigma$ K ($\sigma$ = 10 mK) level to 
ensure only emission is included in the final maps. Additionally, an unmasked spectrum was 
extracted at the position of the QSO sightline, which yields an upper column density limit 
of log $N$(H\,{\sc i}$)=19.53$. 

\end{document}